\input harvmac
\input epsf
\input amssym

\noblackbox

\def\coeff#1#2{\relax{\textstyle {#1 \over #2}}\displaystyle}

 \def\cG{{\cal G}}

 \def\cM{{\cal M}}
 \def\cO{{\cal O}}

\def\bfone{\relax{\rm 1\kern-.35em 1}}
\def\IC{\relax\,\hbox{$\inbar\kern-.3em{\rm C}$}}
\def\ID{\relax{\rm I\kern-.18em D}}
\def\IF{\relax{\rm I\kern-.18em F}}
\def\IH{\relax{\rm I\kern-.18em H}}
\def\II{\relax{\rm I\kern-.17em I}}
\def\IN{\relax{\rm I\kern-.18em N}}
\def\IP{\relax{\rm I\kern-.18em P}}
\def\IQ{\relax\,\hbox{$\inbar\kern-.3em{\rm Q}$}}

\def\IR{\relax{\rm I\kern-.18em R}}
\font\cmss=cmss10 \font\cmsss=cmss10 at 7pt
\def\ZZ{\relax\ifmmode\mathchoice
{\hbox{\cmss Z\kern-.4em Z}}{\hbox{\cmss Z\kern-.4em Z}}
{\lower.9pt\hbox{\cmsss Z\kern-.4em Z}}
{\lower1.2pt\hbox{\cmsss Z\kern-.4em Z}}\else{\cmss Z\kern-.4em
Z}\fi}

\def\ZZ{\Bbb{Z}}
\def\IC{\Bbb{C}}
\def\ID{\Bbb{D}}
\def\IF{\Bbb{F}}
\def\IH{\Bbb{H}}
\def\II{\Bbb{I}}
\def\IN{\Bbb{N}}
\def\IP{\Bbb{P}}
\def\IQ{\Bbb{Q}}
\def\IR{\Bbb{R}}

\def\Nb{\overline{N}}
\def\Pb{\overline{P}}
\def\Qb{\overline{Q}}

%
%
 %
%

\lref\MathurAI{
  S.~D.~Mathur,
``The quantum structure of black holes,''
  arXiv:hep-th/0510180.
}

\lref\MateosQS{
D.~Mateos and P.~K.~Townsend,
``Supertubes,''
Phys.\ Rev.\ Lett.\  {\bf 87}, 011602 (2001)
[arXiv:hep-th/0103030].
}

%
\lref\MathurHJ{
S.~D.~Mathur, A.~Saxena and Y.~K.~Srivastava,
``Constructing 'hair' for the three charge hole,''
Nucl.\ Phys.\ B {\bf 680}, 415 (2004)
[arXiv:hep-th/0311092].
}
%
\lref\MathurSV{
S.~D.~Mathur,
``Where are the states of a black hole?,''
arXiv:hep-th/0401115.
}
%
\lref\GiustoID{
S.~Giusto, S.~D.~Mathur and A.~Saxena,
``Dual geometries for a set of 3-charge microstates,''
arXiv:hep-th/0405017.
}
%
\lref\GiustoIP{
S.~Giusto, S.~D.~Mathur and A.~Saxena,
``3-charge geometries and their CFT duals,''
arXiv:hep-th/0406103.
}
%
\lref\GiustoKJ{
S.~Giusto and S.~D.~Mathur,
``Geometry of D1-D5-P bound states,''
arXiv:hep-th/0409067.
}
\lref\BenaWT{
I.~Bena and P.~Kraus,
``Three charge supertubes and black hole hair,''
Phys.\ Rev.\ D {\bf 70}, 046003 (2004)
[arXiv:hep-th/0402144].
}
\lref\BenaDE{
I.~Bena and N.~P.~Warner,
``One ring to rule them all ... and in the darkness bind them?,''
Adv. Theor. Math. Phys. 9 (2006) 1-35
[arXiv:hep-th/0408106].
}
%
\lref\BenaWV{
  I.~Bena,
  ``Splitting hairs of the three charge black hole,''
  Phys.\ Rev.\ D {\bf 70}, 105018 (2004)
  [arXiv:hep-th/0404073].
}
%
%
\lref\ElvangMJ{
H.~Elvang and R.~Emparan,
``Black rings, supertubes, and a stringy resolution of black hole
non-uniqueness,''
JHEP {\bf 0311}, 035 (2003)
[arXiv:hep-th/0310008].
}
%
\lref\ElvangYY{
H.~Elvang,
``A charged rotating black ring,''
Phys.\ Rev.\ D {\bf 68}, 124016 (2003)
[arXiv:hep-th/0305247].
}
%
\lref\EmparanWN{
R.~Emparan and H.~S.~Reall,
``A rotating black ring in five dimensions,''
Phys.\ Rev.\ Lett.\  {\bf 88}, 101101 (2002)
[arXiv:hep-th/0110260].
}
%
%
\lref\EmparanWY{
R.~Emparan,
``Rotating circular strings, and infinite non-uniqueness 
of black rings,''
JHEP {\bf 0403}, 064 (2004)
[arXiv:hep-th/0402149].
}
%
\lref\GauntlettWH{
  J.~P.~Gauntlett and J.~B.~Gutowski,
  ``Concentric black rings,''
  Phys.\ Rev.\ D {\bf 71}, 025013 (2005)
  [arXiv:hep-th/0408010].
}
%
\lref\SenIN{
A.~Sen,
``Extremal black holes and elementary string states,''
Mod.\ Phys.\ Lett.\ A {\bf 10}, 2081 (1995)
[arXiv:hep-th/9504147].
}
%
\lref\StromingerSH{
A.~Strominger and C.~Vafa,
``Microscopic Origin of the Bekenstein-Hawking Entropy,''
Phys.\ Lett.\ B {\bf 379}, 99 (1996)
[arXiv:hep-th/9601029].
}
%
\lref\GauntlettNW{
J.~P.~Gauntlett, J.~B.~Gutowski, C.~M.~Hull, 
S.~Pakis and H.~S.~Reall,
``All supersymmetric solutions of minimal supergravity 
in five dimensions,''
Class.\ Quant.\ Grav.\  {\bf 20}, 4587 (2003)
[arXiv:hep-th/0209114].
%
}
\lref\GutowskiYV{
J.~B.~Gutowski and H.~S.~Reall,
``General supersymmetric AdS(5) black holes,''
JHEP {\bf 0404}, 048 (2004)
[arXiv:hep-th/0401129].
}
%
\lref\HorowitzJE{
  G.~T.~Horowitz and H.~S.~Reall,
``How hairy can a black ring be?,''
  Class.\ Quant.\ Grav.\  {\bf 22}, 1289 (2005)
  [arXiv:hep-th/0411268].
}
%
\lref\BreckenridgeIS{
J.~C.~Breckenridge, R.~C.~Myers, A.~W.~Peet and C.~Vafa,
``D-branes and spinning black holes,''
Phys.\ Lett.\ B {\bf 391}, 93 (1997)
[arXiv:hep-th/9602065].
}
\lref\TseytlinAS{
A.~A.~Tseytlin,
``Extreme dyonic black holes in string theory,''
Mod.\ Phys.\ Lett.\ A {\bf 11}, 689 (1996)
[arXiv:hep-th/9601177].
}
%

\lref\ElvangDS{H.~Elvang, R.~Emparan, D.~Mateos and H.~S.~Reall,
``Supersymmetric black rings and three-charge supertubes,''
  Phys.\ Rev.\ D {\bf 71}, 024033 (2005)
  [arXiv:hep-th/0408120].
}
\lref\ElvangRT{H.~Elvang, R.~Emparan, D.~Mateos and H.~S.~Reall,
``A supersymmetric black ring,''
  Phys.\ Rev.\ Lett.\  {\bf 93}, 211302 (2004)
  [arXiv:hep-th/0407065].
}

\lref\GauntlettQY{
  J.~P.~Gauntlett and J.~B.~Gutowski,
  ``General concentric black rings,''
  Phys.\ Rev.\ D {\bf 71}, 045002 (2005)
  [arXiv:hep-th/0408122].
}

\lref\ElvangSA{
H.~Elvang, R.~Emparan, D.~Mateos and H.~S.~Reall,
``Supersymmetric 4D rotating black holes from 5D black rings,''
  arXiv:hep-th/0504125.
}
\lref\BenaTK{
  I.~Bena and P.~Kraus,
  ``Microscopic description of black rings in AdS/CFT,''
  JHEP {\bf 0412}, 070 (2004)
  [arXiv:hep-th/0408186].
}
\lref\CyrierHJ{
  M.~Cyrier, M.~Guica, D.~Mateos and A.~Strominger,
  ``Microscopic entropy of the black ring,''
  arXiv:hep-th/0411187.
}

\lref\MaldacenaDE{ J.~M.~Maldacena, A.~Strominger and E.~Witten,
``Black hole entropy in M-theory,''
JHEP {\bf 9712}, 002 (1997)
[arXiv:hep-th/9711053].
}
%
\lref\BenaNI{
  I.~Bena, P.~Kraus and N.~P.~Warner,
  ``Black rings in Taub-NUT,''
  Phys.\ Rev.\ D {\bf 72}, 084019 (2005)
  [arXiv:hep-th/0504142].
}
%
\lref\GaiottoXT{
  D.~Gaiotto, A.~Strominger and X.~Yin,
  ``5D Black Rings and 4D Black Holes,''
  arXiv:hep-th/0504126.
}
\lref\MarolfCX{
  D.~Marolf and A.~Virmani,
  ``A black hole instability in five dimensions?,''
  arXiv:hep-th/0505044.
}

\lref\BenaTD{
  I.~Bena, C.~W.~Wang and N.~P.~Warner,
  ``Black rings with varying charge density,''
  arXiv:hep-th/0411072.
}

\lref\ElvangSA{
  H.~Elvang, R.~Emparan, D.~Mateos and H.~S.~Reall,
  ``Supersymmetric 4D rotating black holes from 5D black rings,''
  JHEP {\bf 0508}, 042 (2005)
  [arXiv:hep-th/0504125].
}

\lref\ReallBH{
  H.~S.~Reall,
  ``Higher dimensional black holes and supersymmetry,''
  Phys.\ Rev.\ D {\bf 68}, 024024 (2003)
  [Erratum-ibid.\ D {\bf 70}, 089902 (2004)]
  [arXiv:hep-th/0211290].
}

\lref\BenaAY{
  I.~Bena and P.~Kraus,
``Microstates of the D1-D5-KK system,''
  Phys.\ Rev.\ D {\bf 72}, 025007 (2005)
  [arXiv:hep-th/0503053].
}

\lref\AldayXJ{
  L.~F.~Alday, J.~de Boer and I.~Messamah,
  ``What is the dual of a dipole?,''
  arXiv:hep-th/0511246.
}

\lref\LuninIZ{
  O.~Lunin, J.~Maldacena and L.~Maoz,
  ``Gravity solutions for the D1-D5 system with angular momentum,''
  arXiv:hep-th/0212210.
}

\lref\LuninJY{
  O.~Lunin and S.~D.~Mathur,
  ``AdS/CFT duality and the black hole information paradox,''
  Nucl.\ Phys.\ B {\bf 623}, 342 (2002)
  [arXiv:hep-th/0109154].
}

\lref\BenaVA{
  I.~Bena and N.~P.~Warner,
  ``Bubbling supertubes and foaming black holes,''
  arXiv:hep-th/0505166.
}

\lref\BerglundVB{
  P.~Berglund, E.~G.~Gimon and T.~S.~Levi,
  ``Supergravity microstates for BPS black holes and black rings,''
  arXiv:hep-th/0505167.
}
%
%
%


\Title{
\vbox{
\hbox{\tt hep-th/0512157}
\hbox{\tt CERN-PH-TH/2005-244 }
}}
{\vbox{\vskip -1.0cm
\centerline{\hbox{Sliding Rings and Spinning Holes}}}}
\vskip -.3cm
\centerline{Iosif~Bena${}^{(1)}$,  Chih-Wei Wang${}^{(2)}$ and
Nicholas P.\ Warner${}^{(2,3)}$}

\bigskip
\centerline{{${}^{(1)}$\it School of Natural Sciences }}
\centerline{{\it Institute for Advanced Study }}
\centerline{{\it Einstein Dr., Princeton, NJ 08540, USA }}
\medskip
\centerline{{${}^{(2)}$\it Department of Physics and Astronomy}}
\centerline{{\it University of Southern California}}
\centerline{{\it Los Angeles, CA 90089-0484, USA}}
\medskip
\centerline{{${}^{(3)}$\it Department of Physics, Theory Division}}
\centerline{{\it CERN, Geneva, Switzerland}}
\medskip

\bigskip
\bigskip

We construct smooth supergravity solutions describing a BPS black
ring with a BPS black hole centered at an arbitrary distance above 
the ring. 
We find that as one moves the black hole
the entropy of the ring remains constant, but the angular momentum
coming from the supergravity fluxes changes. Our solutions also
show that in order to merge a BPS black ring with a BPS black hole one
has to increase one of the angular momenta of the ring, and that
the result of the merger is always a BMPV black hole. We also find a 
class of mergers that are
thermodynamically reversible, and comment on their physics.

\vskip .3in
\Date{\sl {December, 2005}}

\vfill\eject

\newsec{Introduction}

An intriguing feature of BPS black holes in five dimensions is the
fact that their two angular momenta have to be equal
\refs{\BreckenridgeIS,\TseytlinAS}. The recent
conjecture \refs{\BenaWT,\BenaWV} and discovery
\refs{\ElvangRT\BenaDE\ElvangDS\GauntlettWH{--}\GauntlettQY}
of BPS black rings has shown that there also exist objects with unequal 
angular momenta, with horizon topology $S^2 \times S^1$. Since black rings
and black holes are mutually BPS, they can be placed at arbitrary
positions relative to one another  and can even be merged.

It is a very interesting problem to investigate the merging process of
a BPS black ring with a BPS black hole and to determine the
end result.  It is intuitively clear that the merger can either
produce a BPS black hole or a BPS black ring: Pushing a tiny black
ring into a vast black hole, or vice versa, should not change the
horizon topology of the vast object.  This raises an interesting
conundrum when one imagines a small black ring being merged with a
large black hole: When the black ring and the black hole are widely
separated, the two angular momenta of the ring are different from each
other.  Hence, the black hole that one might expect to result from the
merger would, naively, have different angular momenta \BenaWT, and
contradict theorems that claim that BPS five-dimensional black holes
must have equal angular momenta \ReallBH.  
 
Another thought-provoking question is whether one can overspin a BMPV
black hole by dumping into it a black ring with more angular
momentum than a black hole can have. A similar gedanken experiment
involving merging a BMPV black hole and a two-charge supertube 
has been investigated in \BenaWT, and more thoroughly in
\MarolfCX\ 
using the Born-Infeld action of the supertube. While this Born-Infeld
analysis yields very important insights into the merger process (like
the fact that the angular momentum in the plane perpendicular to the
ring depends on the distance between the ring and black hole
\MarolfCX), it is still only perturbative and is limited to two-charge
configurations.  It also does not give too much information about the
back-reaction, the behavior of horizons, or the angular momenta 
coming from fluxes.

It is also interesting to study the process in which a
three-charge supertube (zero-entropy black ring) merges with a black
hole.  If one combines two maximally-spinning, BPS black holes with
charges, $Y^{(i)}$, and angular momenta, $J^{(i)} = (Y^{(i)})^{3/2}$,
where $i=1,2$ labels the black hole, the resulting BMPV black hole
satisfies the strict inequality $(J^{(1)} + J^{(2)} ) < (Y^{(2)} +
Y^{(2)})^{3/2}$.  That is, the merger process is strictly irreversible
in that there is an increase in the horizon area.  We find that the
corresponding process for black rings is ``softer,'' in that there is
a reversible process in which a certain family of three-charge supertubes can be added
to a maximally rotating BPS black hole.  In this process, the entropy
remains constant if the three-charge supertube exactly grazes the horizon.  If
three-charge supertubes are indeed microstates of black holes\foot{This has been proposed in \refs{\BenaWT,\BenaVA}; a more general discussion of this framework can be found in \MathurAI\
and references therein}, then there should be
processes by which they can be added reversibly to another
zero-entropy system.  We indeed find that this is possible, and
determine the precise reversible process. 

In this paper we give a comprehensive analysis of the merger
problem by constructing full supergravity solutions that contain a
(three-charge) BPS black ring and a (three-charge) BPS black hole, where the
black hole is at an arbitrary distance above the center of the ring so
that the solution still has a $U(1)$ isometry, corresponding to
rotation in the plane of the ring.  Our solutions are much more
complicated than the $U(1) \times U(1)$ invariant solutions describing
concentric black rings and black holes
\refs{\GauntlettWH,\BenaDE,\GauntlettQY}
and reduce to those when the black hole is moved to the center of the
ring.

As explained in \refs{\BenaDE,\BenaTK}, and proved in
\refs{\BenaNI,\GaiottoXT}, black-ring solutions have microscopic
charges and angular momenta different from those measured at infinity
because of charges and angular momenta dissolved in fluxes.  We find
that as one changes the location of the black hole, the angular
momentum in the plane of the ring remains fixed, but the other angular
momentum changes. The varying part of the angular momentum precisely
depends on the product of the magnetic charges of the ring and the
electric charges of the black hole. As expected, this component of the
angular momentum increases as the black ring and the black hole approach
one another.
 
If the black hole charges are sufficiently large, then the black ring
and the black hole merge. When this happens, we find that in order to
bring the black ring up to the black hole horizon one must increase
the angular momentum in the plane perpendicular to the ring, so that,
right before the merger, this component of the angular momentum is
exactly equal to the angular momentum in the plane of the ring. Thus,
our analysis shows that the result of such an axially symmetric merger
of a BMPV black hole and a BPS black ring is always another BMPV black
hole.

We have also computed the general expression for the entropy of the
black ring and black hole, and since it depends on eleven parameters,
a general analysis is rather complicated.  We therefore consider a
reduced (five-parameter), but representative, sub-class of black holes
and black rings.  We show that if the black hole has sufficient charge
for the merger to take place, then the resulting black hole has a
total area at least as large as the total horizon area of the
original, widely separated, black hole and black ring.  We therefore
show, at least for this sub-class, that the area-increase theorem is
respected, and therefore that one cannot ``over-spin'' the black hole
by pushing a black ring into it.
 
In section 2 we re-write the $U(1) \times U(1)$ invariant solution
that describes a black ring with a concentric black hole, and we
carefully identify the black-ring microscopic charges and their
relation to the charges of the solution.  In section 3 we solve the
equations that underlie supersymmetric solutions in five dimensions
\refs{\BenaDE,\GauntlettNW,\GutowskiYV}, by using the linear algorithm
discovered in \BenaDE, and we find the exact solution corresponding to
a black hole at an arbitrary distance away from the black ring center.
Those interested only in the solutions can find them summarized in
section 4. While our solutions are constructed in M theory
compactified on $T^6$, they can be trivially extended to any $U(1)^N$
supergravity in five dimensions.  In section 5 we find the charges
and angular momenta of the solutions while, in section 6, we analyze
black-ring black-hole mergers, as well as the possibility of
overspinning a BMPV black hole using black rings.  Finally, in section
7 we present conclusions and suggestions for future research.

\newsec{A black hole at the center of the ring}

As shown in \refs{\GutowskiYV,\BenaDE}, an M-theory background that
preserves the same supersymmetries as three orthogonal M2-branes
can be written as:
\eqn\background{\eqalign{ ds_{11}^2& =  - \left({1 \over Z_1 Z_2
Z_3}\right)^{2/3} (dt+k)^2 + \left( Z_1 Z_2 Z_3\right)^{1/3}
h_{mn}dx^m dx^n \cr &+ \left({Z_2 Z_3 \over
Z_1^2}\right)^{1/3}(dx_1^2+dx_2^2) + \left({Z_1 Z_3 \over
Z_2^2}\right)^{1/3}(dx_3^2+dx_4^2) + \left({Z_1 Z_2 \over
Z_3^2}\right)^{1/3}(dx_5^2+dx_6^2) \,, \cr &   \cr {\cal A} & =
A^1 \wedge dx_1 \wedge dx_2 +A^2 \wedge dx_3 \wedge dx_4 + A^3
\wedge dx_5 \wedge dx_6~,}}
where $A^I$ and $k$ are one-forms in the five-dimensional space
transverse to the $T^6$.  The metric $h_{mn}$ can be any
four-dimensional hyper-K\"ahler metric, but in this paper we
focus on black rings and black holes in $\IR^{(4,1)}$, so we take
this space to be $\IR^4$.
When written in terms of the ``dipole field strengths,''    $\Theta^I$,
\eqn\Thetadefn{\Theta^I \equiv d A^I + d\Big(  {dt +k \over
Z_I}\Big)\,, }
the BPS equations simplify to\foot{Note that the field strengths
used here have a different normalization from those of \BenaDE.}
\refs{\GutowskiYV,\BenaDE}:
\eqn\eom{\eqalign{ \Theta^I  &= \star_4 \Theta^I \cr \nabla^2  Z_I
& = {1 \over 2  } C_{IJK} \star_4 (\Theta^J \wedge \Theta^K) \cr
dk + \star_4 dk &= Z_I \Theta^I~,}}
where $\star_4$ is the Hodge dual on $\IR^4$, and for M theory on $T^6$, 
$C_{IJK} = |\epsilon_{IJK}|$. 
In addition to the M2 branes, this solution contains three sets of M5
branes that wrap four-dimensional tori in the $4567$, $2367$ and 
$2345$ directions respectively, as well as a closed curve in $\IR^4$. 
This curve describes a black-ring profile.

In principle, one can solve these equations to find the solution for an
arbitrary distribution of black holes and black rings.  This is done in
three steps \BenaDE. One first solves for the self-dual field strengths,
$\Theta^I$, sourced by the M5 branes. The second step is to find the 
harmonic functions sourced by the actual M2 branes present in the 
solution, and by the M2 charge coming from the supergravity fields
($C_{IJK} \star_4 (\Theta^J \wedge \Theta^K)$). The third step is to
solve for the angular momentum vector, $k$.

To describe the round black ring it is natural to think of the spatial
$\IR^4$ as $\IR^2 \times \IR^2$ with spherical polar coordinates
$(z,\phi)$ and $(r,\psi)$ in which the metric becomes:
\eqn\firstRmet
{d\vec y \cdot d \vec y ~=~ (dz^2 ~+~ z^2 \, d \psi^2) ~+~
(dr^2 ~+~ r^2 \, d \phi^2)  \,.}
One then locates the ring at $r=0$ and $z=R$. However, it greatly
simplifies calculations  if one introduces a coordinate system 
that makes the dipole fields, $\Theta^I$, very simple. Indeed, it is
conventional to use the coordinates  \ElvangRT:
\eqn\xycoords{
x ~=~  -{ z^2+r^2 - R^2 \over  \sqrt{((z-R)^2 + r^2)( (z+R)^2 + r^2 )}} \,, \qquad
y ~=~ - { z^2+r^2 + R^2 \over  \sqrt{((z-R)^2 + r^2)( (z+R)^2 + r^2 )}} \,,}
in which one has $ -1 \le x \le 1$, $-\infty < y \le -1$.   
In these coordinates, the metric on $\IR^4$ becomes:
\eqn\flatmet{ ds^2_{\IR^4}= {R^2 \over (x-y)^2}\, \bigg( {dy^2 \over y^2 -1}
+ (y^2-1)\, d \psi^2  +{dx^2 \over 1-x^2} + (1-x^2) \, d \phi^2  \bigg)\,.}
In this system the ring is located at $y = -\infty$, while spatial infinity is at  
$x \to -1, y \to -1$ with $(x+1)/(y+1)$ finite.

Since the  black hole does not contain any M5 brane dipole moments, the
$\Theta^J$ are the same as those of a pure ring\foot{In the metric
\flatmet\ we take $\epsilon^{yx\psi\phi} = +1$.}:
\eqn\fluxes{
\Theta^{J} ~=~  2\, q^J \, (d x \wedge d \phi ~-~ d y \wedge d \psi) \,.}

The harmonic functions are also given by the simple superposition of the
ring harmonic functions and the harmonic functions, $H_I$,  of the black hole 
\eqn\concentricz{Z_I ~=~ 1 ~+~ {\Qb_I \over R} (x-y) ~-~{2 \, C_{IJK}\, 
q^J q^K \over  R^2}(x^2 - y^2) ~+~ H_I }
For the present we consider the solution with the black hole at the
center of the ring and so one has \refs{\BenaDE,\GauntlettQY}:
\eqn\Hidefnc{H_I ~=~  - {Y_I \over R^2}\, {x-y \over x+y } \,,}
where $Y_I$ are the charges of the black hole.
For this configuration the angular momentum components are:
\eqn\keq{\eqalign{
k_{\psi} = &   (y^2-1)\,\left({C \over 3} \,(x+y) ~+~
{B \over 2}  - {D \over R^2( x+y)} + {K \over R^2 (x+y)^2 }\right)
~-~  \, A \,(y+1)  \,, \cr
k_{\phi} = &    (x^2-1)\,\left({C \over 3}\, (x+y)~+~
{B \over 2} - {D \over R^2(x+y)} + {K \over R^2 (x+y)^2 } \right)\,.
}}
where $K$ represents the angular momentum of the BMPV black hole and 
\eqn\defs{\eqalign{A & ~\equiv~  2 \big(\sum q^I\big) \,, \qquad \qquad 
B ~\equiv~ {2\over R} \, (Q_I q^I )\,, \qquad  \cr
C & ~\equiv~  -  {8\,  C_{IJK} \, q^I q^J q^K \over R^2}\,, \qquad 
D  ~\equiv~  { 2 } \, Y_I q^I}}
The relation between the quantized ring and black-hole charges and the 
parameters appearing in the solution are:
\eqn\rel{\Qb_I = {\Nb_I \, l_p^6 \over 2 L^4 R}\,,
\qquad q^I={n^I \, l_p^3 \over 4 L^2}\,,
\qquad Y_I = {N^{\rm BH}_I \, l_p^6 \over L^4}\,, 
\qquad  K = { J^{\rm BMPV} \, l_p^6 \over L^6}\, 
.}
where $L$ is the radius of the circles that make up the $T^6$ (so that
$V_6 = (2\pi L)^6$).  The asymptotic charges, $N_I$, of the solution
are the sum of the microscopic charges on the black ring, $\Nb_I$, the charges of the
black hole, $N^{\rm BH}_I$, and the charges dissolved in fluxes:
\eqn\charge{N_I ~=~  \Nb_I ~+~ N^{\rm BH}_I ~+~ \coeff {1}{2} \,
C_{IJK}\,n^J n^K \,.} 
The angular momenta of this solution are:
\eqn\jsol{\eqalign{
J_1 =& J_{\Delta}  ~+~ \Big( \coeff{1}{6}\, C_{IJK} \, n^I n^J n^K  ~+~ 
\coeff{1}{2}\,\Nb_I n^I ~+~ N^{\rm BH}_I n^I  ~+~  J^{\rm BMPV} \Big)\, \cr 
J_2 =& - \Big( \coeff{1}{6}\,C_{IJK} \, n^I n^J n^K  ~+~ 
\coeff{1}{2}\,\Nb_I n^I ~+~ N^{\rm BH}_I n^I ~+~  J^{\rm BMPV} \Big) \,, }}
where
\eqn\jdel{J_{\Delta} \equiv {R^2 L^4 \over l_p^6}\, \Big(\sum n^I \Big)\,.}
The entropy of the ring is:
\eqn\sring{S = {2 \pi A \over \kappa_{11}^2} = \pi \sqrt{\cal M}} 
where 
\eqn\mdef{\eqalign{
{\cal M} ~ \equiv~ &
2\, n^1 n^2 \Nb_1 \Nb_2 +2 \,n^1 n^3 \Nb_1 \Nb_3  
+2\, n^2 n^3 \Nb_2 \Nb_3 -  (n^1 \Nb_1)^2 -  (n^2 \Nb_2)^2 
-  (n^3 \Nb_3)^2  \cr
&-  4\, n^1 n^2 n^3 (J_{\Delta} + n^I N^{\rm BH}_I) \,.
}}
Since the entropy of the black ring is the square root of the
$E_{7(7)}$ quartic invariant of the microscopic charges of the ring
\BenaTK, equation \mdef\ implies that the microscopic angular momentum
of the ring is:
\eqn\jt{J_T ~=~  J_{\Delta} + n^I N^{\rm BH}_I ~=~{R^2 L^4 \over l_p^6}\, 
\Big(\sum n^I \Big) ~+~  n^I N^{\rm BH}_I \,.}
Hence, the angular momenta of the solution may be re-written in terms
of fundamental charges as:
\eqn\jcenter{\eqalign{
J_1 ~=~ & J_{T}  ~+~ \Big( \coeff{1}{6}\,C_{IJK} \, n^I n^J n^K  ~+~  \coeff{1}{2}\, 
\Nb_I n^I  ~+~  J^{\rm BMPV}\Big)\, \cr 
J_2 ~=~ & - \Big( \coeff{1}{6}\, C_{IJK}\,  n^I n^J n^K  ~+~ 
\coeff{1}{2}\, \Nb_I n^I ~+~ N^{\rm BH}_I n^I ~+~  J^{\rm BMPV} \Big) }}
Notice that in this form, $J_1$ contains no contribution coming from
the combined effect of the electric field of the black hole and the
magnetic field of the black ring.  Such a contribution only appears in
$J_2$.

In an adiabatic process in which one moves the black hole into the
center of a black ring one will need to understand how to keep the
black ring (and black hole) ``the same'' during the process.
Obviously the quantized charges $N_I$, $N_I^{BH}$ and $n^I$, which can
be measured on suitable Gaussian surfaces \HorowitzJE, must remain
unchanged.  Furthermore, because of the connection with the
microscopic charges of the four-dimensional black hole and the $E_{7(7)}$
invariant, one might expect $J_T$ to remain unchanged as well.  This
leads one to expect that the only thing that could change is $J_2$,
and only through the term that represents the contribution from the
combined effect of the electric field of the black hole and the
magnetic field of the black ring.  We will show in section 5 that
this intuitive expectation is precisely born out by the exact
solution.

\newsec{Obtaining the new solution}

\subsec{Setting up the problem}

The dipole fluxes, $\Theta^I$, are determined solely by the position
of the ring and so remain unchanged.  The harmonic functions, $Z_I$,
are sourced by the M2 branes and therefore when we move the black hole
off-center we must make a simple translation of the source, \Hidefnc,
that corresponds to the black hole.  The non-trivial consequence of
this lies in the third step of the linear algorithm where one solves
for the angular momentum vector: There is a more complicated
contribution generated by the electric field of the brane interacting
with the magnetic field of the ring.

If we locate a black hole  at $( r, \phi, z,\psi) = (a,0,b,0)$, then the 
Euclidean distance, $d^2$, from a generic point, $( r, \phi, z,\psi)$,  
to the black hole is given by:
\eqn\Eucldist{  d^2  ~=~ (r^2 + a^2 - 2 a r \cos\phi) ~+~
(z^2 + b^2 - 2 b z \cos\psi)  ~=~ - {R^2 \over (x-y)} \, \rho \,}
where
\eqn\rhodefn{ \rho ~\equiv~ (1 + \alpha^2 + \beta^2)\, y ~+~
(1 - \alpha^2 - \beta^2)\, x ~+~  2\,\alpha\, \sqrt{1- x^2}\, \cos \phi ~+~
2\,\beta\, \sqrt{y^2-1}\, \cos \psi\,,}
and $\alpha \equiv {a \over R}$, $\beta \equiv {b \over R}$.  Note
that $\rho \le 0$.  For a single black hole with charges, $Y_I$, we
will orient the coordinates so that $\alpha, \beta \ge 0$, and take:
\eqn\Hidefn{H_I ~=~ - {Y_I \, (x-y) \over  R^2 \, \rho } \,.}
This is simply the translation of \Hidefnc\ to the new center:
$(a,0,b,0)$.  For the present we will make no specific assumptions
about the functions, $H_I$, except that they vanish at infinity and
are regular on the black ring.

The $H_I$ now generate another source term on the right-hand side of
the third equation in \eom.  We therefore define $\hat k$ by:
\eqn\kphi{ k_\psi~=~   (y^2-1)\,\big(\, \coeff{1}{3}\,C \,(x+y) +
\coeff{1}{2} \,B \big) -  A\, (y+1) ~+~ \hat k_\psi  \,, }
\eqn\kpsi { k_\phi~=~    - (1-x^2)\,\big(\, \coeff{1}{3}\,C \,(x+y) +
\coeff{1}{2} \,B \big)   ~+~ \hat k_\phi  \,,}
\eqn\kx{ k_x ~=~   \hat k_x \,, \qquad k_y ~=~   \hat k_y \,,}
where $A,B,$ and $C$ have been defined in \defs, and  $\hat k$ satisfies:
\eqn\hatkeqn{ d\hat k+ *d\hat k ~=~   2 \, q^I \, H_I \,
(d x \wedge d \phi ~-~ d y \wedge d \psi)\,.}
The expression for $k - \hat k$  is simply the angular momentum vector
of the known black-ring solution.

The boundary conditions for $\hat k$ are determined first by requiring
that it is non-singular, except possibly at the black ring and at the
black hole, and then by requiring that there are no closed time-like
curves (CTC's) in the solution.  In practice the latter is hard to
establish globally, and we will not do it here, but there is a
significant danger of getting CTC's near the ring and near the black
hole, and it is by removing these CTC's that we fix the final boundary
conditions on $\hat k$.

As discussed in \BenaDE, adding angular momentum to a black hole  
corresponds to adding a homogeneous solution of the third equation in \eom. This 
homogeneous solution is centered on the black hole, which is now  
away from the center of the coordinate system. By shifting the
homogeneous solution in \keq\ to $r=a$ and re-expressing it in the $x-y $ 
coordinate system we obtain:
\eqn\Jbh{\eqalign{
\hat k^{BMPV}_x~&=~ {\alpha K (x y-1) \sin \phi  
\over R^2 \sqrt{1-x^2} \rho^2}  \,, \qquad   
\hat k^{BMPV}_y~ =~ {\alpha K \sqrt{1-x^2}~ \sin \phi \over 
R^2 \rho^2}  \cr  
\hat k^{BMPV}_\psi~&=~ {K \left(y^2-1\right) \over R^2 \rho ^2} \,, \qquad
\hat k^{BMPV}_\phi~ =~ { K \left[x^2 -1 + \alpha(x-y) \sqrt{1-x^2} ~ 
\cos \phi \right] \over R^2 \rho^2} 
}}
Since the equations determining $k$ are linear, and $ k^{BMPV}$
satisfies \eom\ by construction, we will put it aside for now, and
add it to the final solution at the very end.

Finally, the system the equations for $k$ is gauge invariant.  Note that
a gauge transformation, $k \to k + df$, for some function, $f$, of the
spatial coordinates, is equivalent to the coordinate change, $t \to t
+ f$.  Thus gauge fixing amounts to adjusting the rotational behavior
of the coordinate system.  There are several natural choices for the
gauge, but we will typically choose $k_y =0$ because it simplifies
the analysis of the metric near the ring.   Also note that we may
need to impose some boundary conditions on the gauge choice so
as to make sure that the surfaces of constant time are a non-rotating
frame at infinity.

The explicit system of equations that we have to solve is:
\eqn\ksystem{\eqalign{  (y^2 -1)\,(1-x^2) \, (\partial_y \hat k_x -
\partial_x \hat k_y) ~+~  (  \partial_\psi \hat  k_\phi   -
\partial_\phi \hat  k_\psi)& ~=~ 0 \,, \cr
(y^2 -1)\,(\partial_y \hat k_\phi - \partial_\phi \hat  k_y) ~+~  (1-x^2)\,
(\partial_x \hat k_\psi - \partial_\psi \hat k_x)    & ~=~ 0  \,, \cr
(\partial_y \hat k_\psi ~-~\partial_\psi \hat k_y)
~+~ (\partial_\phi \hat k_x -
\partial_x \hat k_\phi)  &  ~=~ -   2 \, q^I \, H_I \,.}}
%

\subsec{The solution for a vertically displaced black hole }

In this paper we will focus upon the configuration depicted 
in Fig.~1, that is, solutions with  vertically displaced black holes in
which the $U(1)$ symmetry of the plane of the ring is preserved, but the $U(1)$
symmetry in the $(r,\phi)$-plane is broken.  Hence, we will take $b
= \beta =0$ and so \rhodefn\ reduces to:
\eqn\newrhodefn{ \rho ~\equiv~ (1 + \alpha^2)\, y 
~+~ (1 - \alpha^2)\, x ~+~
2\,\alpha\, \sqrt{1- x^2}\, \cos \phi \,.}
The solution we seek must therefore be independent of the angle $\psi$.
If we now go to the gauge with $k_y=0$, one can eliminate between the
equations of  \ksystem\ to show that:
\eqn\secorder{(y^2-1)\,\del_y^2\, \hat k_\psi ~+~ {1 \over (1- x^2)} \,\del_\phi^2\, 
\hat k_\psi  ~+~  \del_x\big((1- x^2)\, \del_x\, \hat k_\psi \big) ~=~  -   2 \, q^I 
\,(y^2-1)\,\del_y H_I \,.}
Using this equation, and the form of the original solution, \keq,
(with $\alpha =0$), it is not very difficult to find a particular
solution for the source term in \secorder.  The more subtle issue is
the careful choice of the homogeneous solution in \secorder.  Indeed,
the following is a homogeneous solution for all values of the constants
$a_j$:
\eqn\homsol{\hat k_\psi^{(0)} =a_1 ~+~ a_2 \, y ~+~ a_3\, \rho^{-1}\,
((1+\alpha^2)\,x  ~+~ (1- \alpha^2)\,y )\,.}
The correct admixture of this with the particular solution is determined
by the boundary conditions in all components of $k$.  
The solution we want is: 
\eqn\kphires{\eqalign{k_\psi~=~  & (y^2-1)\,\big(\, \coeff{1}{3}\,C \,(x+y) +
\coeff{1}{2} \,B \big) -  A\, (y+1) \cr
& ~-~   { D \over 2 \, (1 + \alpha^2)\,R^2} \, (y+1) \Big[  1~-~ {1 \over \rho}\,
 ( (1 + \alpha^2)\,(x-y)  + 2  )     \Big]   \,,}}
%

\goodbreak\midinsert
\vskip .2cm
\hskip -0.8cm
\centerline{ {\epsfxsize 2.7in\epsfbox{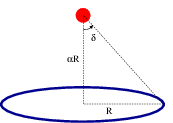}}}
\vskip  .15cm
\leftskip 2pc
\rightskip 2pc\noindent{\ninepoint\sl \baselineskip=8pt
{\bf Fig.~1}:  This shows the configuration of the black hole and black
ring that is described by the new solution. The parameter,  $\alpha$,
is related to the angle of approach, $\delta$, by $\alpha \equiv \cot {\rm \delta}$
}
\endinsert

Given this one can now integrate \ksystem\ to obtain the other components
of $\hat k$.  To express the final result it is useful to define:
\eqn\sigmadefn{ \sigma ~\equiv~ (1 + \alpha^2)  ~+~ (1 - \alpha^2)\, x ~+~
2\,\alpha\, \sqrt{1- x^2}\, \cos \phi  }
and introduce the  function:
\eqn\Fdefn{F(x,y) ~\equiv~   ((1 + \alpha^2 )\,x  +  (1 - \alpha^2) )\,
\Big({1 \over \sigma \, \rho} ~+~ {1 \over \sigma^2}\,
\log\Big(1 - {\sigma \over \rho}\Big) \Big)\,.}
Then one has
\eqn\kpsires{\eqalign{k_\phi~=~  & - (1-x^2)\,\big(\, \coeff{1}{3}\,C  
\,(x+y) +
\coeff{1}{2} \,B \big)  ~-~   { D \, (1- x^2)  
\over 2 \,   R^2} \,
{1 \over \sigma}\, \log  \Big(1 - {\sigma \over \rho} \Big)  \cr
&  ~+~ { D \, (1- x^2) \over 2 \, (1 + \alpha^2)  \, R^2}   \Big((1  
- \alpha^2) -
{2\, \alpha \,  x\,  \cos \phi \over \sqrt{1 - x^2}} \Big)
\, \Big( F(x,y) ~+~ {1 \over \rho} \Big) \cr
&~+~   { D\,  \log(1 + \alpha^2) \over 2\, \alpha \, (1 + \alpha^2) 
\, R^2} \,
   \sqrt{1- x^2}  \cos \phi    ~+~   { D\, (\alpha^2 - \log(1 +  
\alpha^2)) \over
2\, \alpha^2 \, (1 + \alpha^2)\, R^2} \, (1-x^2)\, \cos 2\phi  \,,}}
\eqn\kxres{\eqalign{k_x ~=~  & {\alpha\, D\, \sin\phi  \over  (1 +  
\alpha^2)\, R^2\,
\sqrt{1- x^2}} \, \Big( {1 \over \rho} ~+~F(x,y) \Big)~-~   { D\,   
\log(1 + \alpha^2)
\over 2\, \alpha \, (1 + \alpha^2)\, R^2} \, {x  \over   \sqrt{1-  
x^2}} \sin\phi \cr
&~-~   { D\, (\alpha^2 - \log(1 + \alpha^2)) \over
2\, \alpha^2 \, (1 + \alpha^2)\, R^2} \, x \sin 2\phi   \,,}}
In integrating to obtain $\hat k$ there are some important integration
``constants'' and gauge ambiguities to be resolved.   This is done  
first by
making sure that  angular  momentum vector, $k$, is regular  
everywhere except at
the black hole ($\rho =0$) and at the black ring ($y = -\infty$).
The last two terms in $k_\phi$ and in $k_x$ are,
in fact, pure gauge and could, in principle, be discarded.  Gauge
transformations in $k$ do, however, amount to the choice of the  
spatial sections
of the metric (the surfaces of constant $t$) and the inclusion of  
these particular gauge
terms in $k$ amounts to the choice of a non-rotating coordinate system
at infinity.

\subsec{Regularity}

Looking at the form of $k$, it appears that there might be 
singularities at $\sigma =0$, but if one expands the logarithms for  
small $\sigma$ one easily sees that $k$ is smooth at $\sigma =0$.  The  
logarithms may also be re-written as $\log((1 + \alpha^2)(y-1)/\rho)$ and, away  
from the black hole and black ring, the argument is positive 
definite because $\rho$ is negative definite and $-\infty < y < -1$.

The angular momentum component, $k_x$, is singular at $x = \pm1$, 
both in \Jbh\  and in the new solution, but this singularity  
is a coordinate
artifact because the $(x, \phi)$ coordinate system is degenerate at
these points.  Obviously, $\hat k^{BMPV}$ cannot be singular at 
$x = \pm 1$ because it is just the translation of a regular vector 
field to a different coordinate system

To show that $k$ in is non-singular at $x = \pm 1$, one  
first collects all the terms involving $ \sqrt{1- x^2}$:
\eqn\ksqrt{\eqalign{\tilde k ~=~ & { \alpha\,  D  \over  (1 +  
\alpha^2)  \, R^2}  \,
\Big({\sin \phi \over \sqrt{1 - x^2}  } \, dx - x\, \sqrt{1 - x^2}   
\cos \phi \, d\phi \Big) \,
\Big( F(x,y) ~+~ {1 \over \rho} \Big) \cr
&\qquad ~+~d\,\Big( { D\,  \log(1 + \alpha^2)
\over 2\, \alpha \, (1 + \alpha^2)\, R^2} \,  \sqrt{1- x^2} \, \sin 
\phi  \Big)\,.}}
Now introduce  the Cartesian coordinates: $u = \sqrt{1- x^2} \cos \phi$,
$v = \sqrt{1- x^2} \sin \phi$, which are regular near $x = \pm 1$.   
Indeed,
one has:
\eqn\cartmet{ du^2 ~+~ dv^2 ~=~  {x^2 \, dx^2 \over (1 - x^2)} ~+~
(1 - x^2) \, d \phi^2 \,,}
which is conformal to the $(x,\phi)$ part of \flatmet\ near  
$x= \pm 1$.
Now observe that:
\eqn\diffform{\eqalign{\tilde k  ~=~  & - { \alpha\,  D  \over  (1  
+ \alpha^2)  \, R^2}  \,
\Big( F(x,y) ~+~ {1 \over \rho} \Big)  \,  x\, dv ~+~d\,\Big( { D 
\,  \log(1 + \alpha^2)
\over 2\, \alpha \, (1 + \alpha^2)\, R^2} \,  v  \Big)\cr
&~+~ { \alpha\,  D  \over  (1 + \alpha^2)  \, R^2}   \,
   \Big( F(x,y) ~+~ {1 \over \rho} \Big)    \,  v\, dx   \,,}}
which means that $k$ is completely regular at $u=v=0$, or $x = \pm 1$.
One can also do a gauge transformation that makes $k$ manifestly regular 
at  $x = \pm 1$. This is done in the Appendix.

\subsec{Horizons}

The analysis of the horizon of a black hole is almost trivial.  
Near the black hole,
the functions, $Z_I$, and the warp factor in \background\ behave as:
\eqn\BHasymp{Z_I ~\sim~ {Y_I \over d^2} \qquad \Rightarrow  \qquad
\left( Z_1 Z_2 Z_3\right)^{1/3} \sim {\left( Y_1 Y_2 Y_3\right)^{1/3} \over d^2}\,,} 
Thus, for a non-rotating black hole, the three-spheres around the
black hole (at $d=0$) limit to a sphere of constant radius of order 
$(Y_1 Y_2 Y_3)^{1/3}$. (If the black hole rotates then there are
terms that contribute at the same order coming from the angular
momentum vector, $\hat k^{BMPV}$.)
The terms in $Z_I$ that are  relevant for the horizon geometry 
are all $\cO(d^{-2})$, while nearby BPS
objects only modify the $Z_I$ at terms of $\cO(d^{0})$ as $d \to 0$.
Therefore, the horizon geometry of the black hole is
completely oblivious to any other BPS objects nearby.
 
 Conversely, if one carefully examines the derivation of the near-ring metric 
one sees that leading powers of $y$  in $k_\psi$ are 
determined  by requiring them to cancel against the {\it leading and first sub-leading} 
orders of divergence in the warp-factor, $ (Z_1 Z_2 Z_3)^{1/3}$.  (Remember that 
the ring is located at $y = -\infty$.)  Terms at the second sub-leading order then
generate {\it finite} corrections to the near-ring geometry.  Thus constant terms
in the $Z_I$ and terms of order $y$ in $k_\psi$ can only make a finite contribution 
to the near horizon geometry.   This means that the nearby black hole 
will not cause any new singular behavior near the ring horizon, but    
the horizon geometry can, and indeed does, have a finite
response to such nearby BPS objects.  The horizon area of the
deformed black ring will be given in section 5.

The fact that the entropy density depends on the value of the black-hole 
electric fields at the horizon, means that a
black-hole/black-ring solution in which the $U(1)$ along the ring is not
preserved will have a ring horizon whose area changes as one goes along
the ring. Such configurations are very instructive to study, and are
currently under examination.  Here, however, we focus on the processes
that preserve the $U(1)$ symmetry of the ring.

\newsec{The complete solution}

The metric and forms are given by equations \background\ 
and \Thetadefn, with
\eqn\sfluxes{
\Theta^{J} ~=~  2\, q^J \, (d x \wedge d \phi ~-~ d y \wedge d \psi) \,,}
and
\eqn\sconcentricz{Z_I ~=~ 1 ~+~ {\Qb_I \over R} (x-y) 
~-~{2 \, C_{IJK}\, 
q^J q^K \over  R^2}(x^2 - y^2)  - {Y_I \, (x-y) \over  R^2 \, \rho } 
\,, }
where 
\eqn\srhodefn{ \rho = (1 + \alpha^2 )\, y ~+~
(1 - \alpha^2)\, x ~+~  2\,\alpha\, \sqrt{1- x^2}\, \cos \phi,}
and $\alpha \equiv {a \over R}$. To express the angular momentum
vectors we introduce the quantities
\eqn\ssigmadefn{\eqalign{ \sigma &~\equiv~ (1 + \alpha^2)  ~+~ (1 - \alpha^2)\, x ~+~
2\,\alpha\, \sqrt{1- x^2}\, \cos \phi \, ,\cr
F(x,y) &~\equiv~   ((1 + \alpha^2 )\,x  +  (1 - \alpha^2) )\,
\Big({1 \over \sigma \, \rho} ~+~ {1 \over \sigma^2}\,
\log\Big(1 - {\sigma \over \rho}\Big) \Big)\,.}}
The total angular momentum vector is obtained by adding the solution 
obtained in the previous section and the homogeneous solution 
corresponding to the rotation of the black hole \Jbh. One obtains:
\eqn\skphires{\eqalign{k_\psi~=~  & (y^2-1)\,
\big(\, \coeff{1}{3}\,C \,(x+y) +
\coeff{1}{2} \,B \big) -  A\, (y+1) \cr
& ~-~ { D \over 2 \, (1 + \alpha^2)\,R^2} \, (y+1) \Big[ 1~-~ {1
 \over \rho}\, ( (1 + \alpha^2)\,(x-y) + 2 ) \Big]  ~+~ 
{K \left(y^2-1\right) \over R^2 \rho ^2}\, , 
\cr
%
%
k_\phi~=~  & - (1-x^2)\,\big(\, \coeff{1}{3}\,C  
\,(x+y) +
\coeff{1}{2} \,B \big)  ~-~   { D \, (1- x^2)  
\over 2 \,   R^2} \,
{1 \over \sigma}\, \log \Big(1 - {\sigma \over \rho} \Big) \cr & ~+~ 
{D \, (1- x^2) \over 2 \, (1 + \alpha^2) \, R^2} \Big((1 - \alpha^2)
- {2\, \alpha \, x\, \cos \phi \over \sqrt{1 - x^2}} \Big)
\, \Big( F(x,y) ~+~ {1 \over \rho} \Big) \cr
&~+~   { D\,  \log(1 + \alpha^2) \over 2\, \alpha \, (1 + \alpha^2) 
\, R^2} \,
   \sqrt{1- x^2}  \cos \phi    ~+~   { D\, (\alpha^2 - \log(1 +  
\alpha^2)) \over
2\, \alpha^2 \, (1 + \alpha^2)\, R^2} \, (1-x^2)\, \cos 2\phi  \, \cr
&~+~ { K \left[x^2 -1 + \alpha(x-y) \sqrt{1-x^2} ~ \cos \phi \right] 
\over R^2 \rho^2} \,,
\cr
%
%
k_x ~=~  & {\alpha\, D\, \sin\phi  \over  (1 +  
\alpha^2)\, R^2\,
\sqrt{1- x^2}} \, \Big( {1 \over \rho} ~+~F(x,y) \Big)~-~   { D\,   
\log(1 + \alpha^2)
\over 2\, \alpha \, (1 + \alpha^2)\, R^2} \, {x  \over   \sqrt{1-  
x^2}} \sin\phi \cr
&~-~   { D\, (\alpha^2 - \log(1 + \alpha^2)) \over
2\, \alpha^2 \, (1 + \alpha^2)\, R^2} \, x \sin 2\phi ~+~ 
{\alpha K (x y-1) \sin \phi  
\over R^2 \sqrt{1-x^2} \rho^2}   \,,
\cr
%
%
k_y ~=~ & {\alpha K \sqrt{1-x^2}~ \sin \phi \over 
R^2 \rho^2}~.    
}}
 
The quantities $A,B,C,D$ are defined in \defs, and the relations
between the black ring and black hole microscopic charges and the
parameters appearing in the solution are given in \rel.

\newsec{The charges of the new solution}

The asymptotic charges of the new solutions are the same as those of 
the solution with the black hole in the center of the ring:
\eqn\chargenew{N_I ~=~ \Nb_I ~+~  N^{\rm BH}_I ~+~ \coeff{1}{2} \,
C_{IJK}\, n^J n^K \,.} 
The angular momenta of the solution can easily be obtained by expanding
\kphires\ and \kpsires\ near spatial infinity.  One finds:
\eqn\jsolnew{\eqalign{
J_1 =& J_{\Delta}\,
 ~+~ \Big( \coeff{1}{6} \, C_{IJK} n^I n^J n^K  ~+~ 
 \coeff{1}{2} \, \Nb_I n^I ~+~ {N^{\rm BH}_I n^I \over {1+\alpha^2}}
~+~  J^{\rm BMPV} \Big)\, \cr 
J_2 =& - \Big(  \coeff{1}{6} \, C_{IJK}  n^I n^J n^K  ~+~ \coeff{1}{2} \,  
\Nb_I n^I ~+~ {N^{\rm BH}_I n^I \over {1+\alpha^2}}~+~  J^{\rm BMPV} \Big) }}
where $J_{\Delta}$ is the same as in the concentric configuration:
\eqn\jdelnew{J_{\Delta} \equiv {R^2 L^4 \over l_p^6}\, \Big(\sum n^I \Big)\,.}

One can also find the horizon area of the ring, and read off its entropy:
\eqn\sringnew{S = {2 \pi A \over \kappa_{11}^2} = \pi \sqrt{\cal M} \,,} 
where now
\eqn\mdefnew{\eqalign{
{\cal M}  ~\equiv~&
2\,  n^1  n^2 \, \Nb_1  \Nb_2 + 2\,  n^1 n^3\,  \Nb_1  \Nb_3   +  2 \, n^2  n^3\,  
\Nb_2  \Nb_3 -   (n^1\,  \Nb_1)^2 -  (n^2 \, \Nb_2)^2  -  (n^3\,  \Nb_3)^2  \cr
&-  4 n^1   n^2\  n^3\,  \Big(J_{\Delta}~+~  
{n^I\,  N^{\rm BH}_I \over {1+\alpha^2}} \Big) \,.
}}
As we explained in section 3, the presence of terms proportional to 
$ N^{\rm BH}_I$ in ${\cal M}$ 
indicates that the horizon of the black ring ``feels'' the presence of 
the black hole. In contrast, the black hole horizon is completely insensitive
to the presence of other BPS objects nearby.
 
Again, since the entropy of the black ring is the square root of the $E_{7(7)}$ quartic 
invariant of the microscopic charges of the ring \BenaTK, equation \mdefnew\
implies that the microscopic angular momentum of the ring is:
\eqn\jtnew{J_T ~=~  J_{\Delta} ~+~ {n^I N^{\rm BH}_I \over {1+\alpha^2}}
~\equiv~  {R^2 L^4 \over l_p^6}\, \Big(\sum n^I \Big) ~+~ {n^I N^{\rm BH}_I 
\over {1+\alpha^2}}\,.}
When written in terms of the microscopic charges of the ring, the angular 
momenta of the solution become:
\eqn\jcenternew{\eqalign{
J_1 =& J_{T} ~+~ \Big(  \coeff{1}{6} \, C_{IJK} n^I n^J n^K  ~+~  
\coeff{1}{2} \,  \Nb_I n^I  ~+~  J^{\rm BMPV} \Big)\, \cr 
J_2 =& - \Big(  \coeff{1}{6} \, C_{IJK} n^I n^J n^K  ~+~  \coeff{1}{2} \, 
\Nb_I n^I ~+~ {N^{\rm BH}_I n^I  \over {1+\alpha^2}} ~+~  J^{\rm BMPV} \Big)\,. }}

Our solutions describe black rings with arbitrary charges, dipole
charges and angular momenta, and black holes with arbitrary
charges and angular momenta. However, if we want to study the 
adiabatic merger of a black
hole and a black ring, we have to focus on a subset of our solutions
that describe {\it the same} black hole and {\it the same} black ring
at arbitrary separations.

This implies that the charges $N^{\rm BH}_I $ of the black hole and
the charges $\Nb_I$ and dipole charges $ n^I$ of the black ring must
remain the same. The other quantity that must remain invariant is
$J_{T}$, both because it is a microscopic charge of the black ring,
and because in an adiabatic process the total area of all horizons
must remain constant. Since the black-hole horizon remains unchanged,
the ring horizon area, \mdefnew, must therefore remain constant.  The fact
that $\Nb_I$ and $ n^I$ are the same then automatically implies that in an
adiabatic process $J_T$ does not change.

This implies that the angular momentum, $J_1$, also remains constant as
one moves the black ring and black hole by adiabatically varying
$\alpha$.  This is to be expected since the $\psi$ translations
remain a symmetry of the solution throughout this process and so
$J_1$ must be conserved.

On the other hand,  the angular momentum component, $J_2$, depends on
the relative position of the ring and of the black hole.  The
relevant term, ${N^{\rm BH}_I n^I \over {1+\alpha^2}}$, comes
from the Poynting vector sourced by the magnetic fields of the
ring and the electric fields of the black hole. When the black ring is
infinitely far away, one has $\alpha \rightarrow \infty$, and this term
vanishes.

It is possible to see intuitively why the angular momentum coming from
fluxes changes as the black hole and the ring move apart. In four
dimensions both magnetic and electric charges are point-like, and the
angular momentum coming from having an electron and a magnetic monopole  
is the same regardless of their relative position.  Similarly,
in five dimensions the angular momentum coming from the magnetic field
of an infinitely long black string and the electric field of a black
hole should be constant. However, if we have a black {\it ring} and a
black hole, the magnetic field of the former decays faster than that
of an infinitely long string, and hence the angular momentum coming
from the integral of $\vec E \times \vec B$ should depend on the
distance between the two objects.
 
The constancy of $J_T$ along with \jtnew\ imply that as the black hole
is brought near the black ring, the embedding radius of the latter,
$R$, must change according to:
\eqn\radius{    R^2 ~=~{ l_p^6  \over L^4 }\,
\,\big(\sum n^I \big )^{-1} \,  \Big(J_T - {n^I N^{\rm BH}_I \over 
{1+\alpha^2}}\Big) \, .}
For fixed microscopic charges this formula gives the radius of the 
ring as a function of the parameter $\alpha$. 

If one of the charges and two of the dipole charges of the black ring
are set to zero, it becomes a two-charge supertube \MateosQS. The physics of a
supertube probe in a BMPV black hole background has been studied
before using the Born-Infeld action of a supertube
\refs{\BenaWT,\MarolfCX}. The supertube analysis gives a nontrivial
check to two of the the phenomena we observe. The first is that one
cannot move the black hole off the center of the ring without changing
$J_2$ \MarolfCX. The second is that the supertube radius changes as
the black hole is moved. In the supertube limit ($\Nb_3,n^1,n^2
\rightarrow 0$) the radius formula \radius\ reproduces exactly the
radius of the two charge supertube in the three-charge black hole
background computed in \refs{\BenaWT,\MarolfCX}.

\newsec{Mergers and acquisitions}

We have seen that as we vary the separation of the black hole and the
black ring, the embedding radius of the ring changes according to
\radius. Before we begin a detailed discussion of this, we want to
underline that even though the embedding radius varies, the physical
size of the ring, as determined by that of the horizon of the black
ring, will remain constant.  Thus requiring the process to be both
adiabatic and retain the $U(1)$ symmetry of the ring means that the
ring will remain rigid.

\subsec{The merging process}

Equation \radius\  shows that for certain values of the charges,
the ring radius, $R$, goes to zero at a non-zero value of $\alpha$.
One should remember that the distance (in $\IR^4$) between the black
hole and the plane of the ring is $\alpha R$, and so the limit $R \to
0$ not only means that the embedding radius is vanishing but also that
the black hole is limiting to the ring plane.  Hence, when $R
\to 0$  the black ring is merging with the black hole.  The
value of $\alpha$ where this happens simply determines the black-hole
latitude at which the ring arrives, or ``crowns,'' the black hole.
(See Fig.~2.)

The physics here is similar to what happens to supertubes and black
rings in Taub-NUT \refs{\BenaAY,\ElvangSA,\GaiottoXT,\BenaNI} as one
changes the Taub-NUT radius. The physical size of the ring is completely 
fixed by its charges. If the size of the space is modified (either by bringing in a
black hole, or by modifying the Taub-NUT radius) the ring migrates to a different location, 
so as to maintain its actual physical size. 

Even though the black hole appears
point-like in the embedding space, it has a physical size determined
by its charges.  
Indeed, it is worth recalling that, in  $\IR^4$ polar coordinates
centered on the black hole (with $r= d \sin \chi$, $z= d \cos \chi$),
the horizon metric of the black hole is:
\eqn\BHhormet{\eqalign{ds^{BH}_3 ~=~ (Y_1 \, Y_2 \, Y_3 )^{1/3} \, 
\Big(d \chi^2 & ~+~ \sin^2 \chi \, d \phi^2 ~+~ \cos^2 \chi \, d \psi^2 \cr 
& ~-~  {K^2 \over Y_1 \, Y_2 \, Y_3  } \,
 (\sin^2 \chi \, d \phi ~-~ \cos^2 \chi \, d \psi)^2\Big) \,.}}
%

\goodbreak\midinsert
\vskip .2cm
\hskip -0.8cm
\centerline{ {\epsfxsize 1.8in\epsfbox{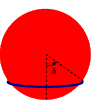}}}
\vskip  .15cm
\leftskip 2pc
\rightskip 2pc\noindent{\ninepoint\sl \baselineskip=8pt
{\bf Fig.~2}:  This shows the black hole and the black ring at the merger. 
The ring has $R=0$. The value of $\alpha$ at the merger gives the 
latitude at which the ring crosses the horizon ($\alpha \equiv \cot {\rm \delta}$).
}
\endinsert

As one can see from \radius, whether a ring and a black hole merge or not, is 
completely independent of the black hole angular momentum.  This 
is perhaps somewhat surprising, particularly given the fact that a 
maximally-spinning black hole has zero horizon area; however one
can see from \BHhormet\ that the circles in the $\psi$-direction, over which 
the ring must slip, have a radius that is bounded below 
(at $\chi = {\pi \over 4}$) by  ${1 \over 2} (Y_1 \, Y_2 \, Y_3 )^{1/6} $, which
is independent of $K$.    This observation is, by no means, a complete
explanation, particularly when the ring approaches the black hole from very
close to the equatorial plane.  It would be interesting to see if one
can understand \radius\ entirely from the geometry of ``black hole
hoopla.''

If the charges are
such that $R$ never becomes zero as $\alpha$ goes to zero, then, as
one passes the black hole through the center of the black ring, the embedding
radius just shrinks to a minimal size, and then re-expands to its
``normal'' radius when the black hole moves infinitely far away.

Equation \jcenternew\ implies that in order to bring the black ring to
the black-hole horizon, one has to increase the angular momentum
$J_2$. For the merger to happen one must have $R \to 0$ for
some value of $\alpha$.  This means that  $J_{\Delta} = 0$, and from
\jsolnew\ we see that  this means $J_1= J_2$.  Thus
 the black object that results from the merger has 
charges, $N_I$, given in \chargenew\ and angular momenta both equal to:
\eqn\jbmpvfinal{J^{\rm BMPV}_{\rm final} =  J_{T}\, ~+~  \coeff{1}{6} \, C_{IJK}\,
 n^I n^J n^K  ~+~ \coeff{1}{2} \,  \Nb_I n^I ~+~  J^{\rm BMPV}\,,} 
and is nothing but a BMPV black hole. Hence, we have shown that to
merge a black ring with a three-charge black hole in this axially
symmetric manner, one has to add enough angular momentum to the
solution so that the resulting black hole has equal angular momenta.

This result solves an old puzzle. When probing supersymmetric black
holes with supertubes, it was found that a supertube can be easily
dumped into a black hole \refs{\BenaWT,\MarolfCX}. Since supertubes
have generically unequal angular momenta, the resulting supersymmetric
black hole would have naively had unequal angular momenta. However,
such supersymmetric black holes were argued not to exist at all in
five-dimensional minimal supergravity \ReallBH. The key to the
solution of this puzzle is the fact that the total angular momentum coming from
fluxes changes as the
supertube moves towards the black hole. When the two merge, the
angular momentum coming from fluxes exactly equals the supertube/ring
angular momentum, and the resulting object has equal angular momenta.

\subsec{Chronology protection and the entropy of mergers}

As discussed above, whether a ring and a black hole merge or not,
is completely independent of the angular momentum of the black
hole. Thus, one can consider a black hole that is maximally
spinning and merge it with a black ring. One can also give the
black ring a very high angular momentum, by choosing a $J_T$ that
renders the ring entropy very close to zero.

The resulting object will have a rather large angular momentum, and 
it is interesting to see if this angular momentum can be larger than 
the maximal allowed angular momentum of the resulting black hole, 
$J_{\rm max} = \sqrt{N_1 N_2 N_3}$. A related question is whether the 
entropy of the black hole that results from the merger is larger than 
the sum of the entropies of the ring and hole that merge. Since the 
entropy of the BMPV black hole is
\eqn\sbmpv{S_{BH} ~=~ 2 \pi \sqrt{N_1 N_2 N_3 - J^2}}
an entropy increase always implies chronology protection, but the 
reverse is not necessarily true.

The general problem of finding whether black hole  mergers with black rings
are thermodynamically irreversible involves eleven parameters: $n^I$,
$N_I$, $N_I^{BH}$, $R$ and $J^{BMPV}_{\rm initial}$, and it is rather too
involved to analyze completely here.  To simplify the algebra we
consider a representative five-parameter sub-class of solutions in
which we set:
\eqn\simpcharges{n^I ~=~ n \,, \qquad \Nb_I ~=~ \Nb \,, \qquad N_I^{BH} ~=~ 
N  \,,   \qquad I=1,2,3 \,.}
We consider a ring starting from infinity with $R = R_\infty$ and
being adiabatically merged with a BMPV black hole of initial
angular momentum, $ J^{BMPV}_{\rm initial}=J$.

The final state must be a BMPV black hole with charges given by \chargenew:
\eqn\Mfinal{ N_{\rm final} ~=~   N ~+~  \Nb ~+~  n^2 \,,}
and with final angular momenta given by \jbmpvfinal:
\eqn\jfinal{J_{\rm final} =   3 \, ( L^4 \, l_p^{-6})\, R_\infty^2 \, n  ~+~  n^3  ~+~ 
\coeff{3}{2} \,   \Nb \,  n  ~+~  J \,.}  
The change of entropy in the merger process is then:
\eqn\DeltaS{\Delta S ~=~ 2\,\pi \sqrt{ N _{\rm final} ^3 - 
J_{\rm final} ^2}  ~-~ \Big( \pi \sqrt{\cM} + 2\,\pi \sqrt{N^3 - J^2}  \Big) \,,}
where $\cM$ is given by \mdef.  To show that this is non-negative,
it is equivalent to show that the function:
\eqn\cGdefn{\cG(n, \Nb, N,  R_\infty, J) ~\equiv~ 
(N _{\rm final} ^3 -  J_{\rm final} ^2) ~-~
 \bigg(\coeff{1}{2} \sqrt{\cM} +   \sqrt{ N^3 - J^2}  \bigg)^2 }
is non-negative.

We will, of course, require that the initial states have non-negative
horizon areas:
\eqn\ringhor{\cM ~=~ 3\,n^2 \Big( \Nb^2 ~-~ 4\,n^2\,  ( L^4 \, l_p^{-6})\, 
R_\infty^2 \Big)  ~\ge~ 0 \quad \Leftrightarrow \quad 
 \Nb ~\ge~ 2\,n \,  ( L^2 \, l_p^{-3})\, 
R_\infty   \,,}
\eqn\BMPVhor{ N^3 - J^2 ~\ge~ 0   \,.}
There is also the condition that the merger actually happens.  We know that
$J_T$ is constant and so 
\eqn\jtequality{J_T ~=~  {R^2 L^4 \over l_p^6}\, \Big(\sum n^I \Big) ~+~ 
{n^I N^{\rm BH}_I  \over {1+\alpha^2}} ~=~   {R_\infty^2 L^4 \over l_p^6}\, 
\Big(\sum n^I \Big)\,.}
Therefore, in order for $R \to 0$ at some value of $\alpha$ one must have:
\eqn\mergercond{( L^4 \, l_p^{-6})\,  R_\infty^2  ~\le~   N   \,.}
Therefore our task is to show that $\cG$ is non-negative in the domain
defined by \ringhor, \BMPVhor\ and \mergercond.

First consider the dependence on $J$.  One can easily check that
${d \cG \over d J} < 0$ at $J=0$ and ${d \cG \over d J} \to + \infty $ as
$J  \to  N^{3/2}$, and so, perhaps rather surprisingly, $\cG$,
is not minimized at $J=   N^{3/2}$.  The actual minimum, as
a function of $J$ occurs at:
\eqn\Jmin{ J~=~  { N^{3/2}\, \big(n^2 + \coeff{3}{2}  \,  \Nb + 
3\,( L^4 \, l_p^{-6})\,  R_\infty^2 \big) \over \sqrt{ \big(n^2 + \coeff{3}{2}  \,  \Nb + 
3\,( L^4 \, l_p^{-6})\,  R_\infty^2 \big)^2 +  \coeff{3}{4}\, \big( \Nb^2 - 4 \, n^2\,  
( L^4 \, l_p^{-6})\,  R_\infty^2 \big) }} \,,}
which clearly lies in the range $0< J \le  N^{3/2}$ and hits
the upper bound if and only if the black ring is actually a supertube
with $\cM=0$.

Now let $\cG_1$ be the function $\cG$ evaluated at the minimizing
value of $J$ in \Jmin.   Furthermore, consider $\cG_1$ as a function
of $R_\infty$.  One can easily check that ${d \cG_1 \over d R_\infty} < 0$
and so the minimum of $\cG_1$ occurs at the maximum value of
$R_\infty$.  We have two bounds on $R_\infty$ given by \ringhor\ and
\mergercond.  The former is the relevant limit if $ N \ge 
{ \Nb^2 \over 4 n^2}$
and the latter is relevant for $  N \le { \Nb^2 \over 4 n^2}$.

Suppose that the bound in \mergercond\ is saturated; then, from  \ringhor, 
one has $ \Nb^2  \ge 4 n^2   N$ and one can also show that:
\eqn\cGonea{\cG_1 ~=~ p_2 ~-~ \sqrt{p_2^2 ~-~ \big( \Nb^2 - 4 \, n^2 \,  
N \big) \, p_1^2} \,,}
where
\eqn\ponetwodefn{p_1 ~\equiv~  \Nb^2 + 3 \, N\, \Nb  + 3 \, 
 N^2 - n^2\,   N \,, \qquad p_2 ~\equiv~  \Nb \, p_1 -
2\, n^2 \,   N (3 N +  \Nb) \,.}
Since $ \Nb^2  \ge 4 n^2  N$, one has $p_1 >0$, and so
$\cG_1 \ge 0$ with equality if and only if $ \Nb^2 = 4 n^2  N$.
That is, $\cG_1 \ge  0$ with $\cG_1 = 0$ if and only if {\it both} bounds, 
 \mergercond\  and  \ringhor, are saturated.
 
Now suppose that the bound in  \ringhor\ is saturated and
introduce the variable $\nu = \sqrt{ N}$.  From \mergercond,  
one has $ \Nb  \le 2 n \nu $.  
One can now show that:
\eqn\cGonea{\eqalign{\cG_1 ~=~ & {1 \over 16 \, n^2}\,(2\, n\, \nu -  \Nb) \, 
\big( 9\,  \Nb^3 + 18\, n\,  \Nb^2 \, \nu  +12 \,n^4 \,  \Nb + 20 \, n^2\,  \Nb^2 \cr 
& \quad + 8\,n^3 \, \nu\, (3\,n^2 - 2 n\, \nu + 3\, \nu^2)  +  
4\,n \, \nu\, \Nb\, (10\,n^2 - 3 n\, \nu + 6\, \nu^2)\big) \,.}}
Note that the quadratic forms that appear as coefficients of  
$ 8 n^3 \nu$ and $4 n \nu$ are both strictly positive.  Thus, in this
limit we also have  $\cG_1 \ge  0$ with $\cG_1 = 0$ if and only if 
{\it both} bounds,  \mergercond\  and  \ringhor, are saturated.

Conversely, suppose that both bounds, \mergercond\  and  \ringhor,
are saturated, then one finds:
\eqn\cGsat{\cG ~=~ {1 \over2\, n}\, (N^{3/2} - J)\, 
(4\, n^4 + 6\, n^2 \, \Nb + 3 \, \Nb^2) \,.}
It follows that the entropy increase is strictly positive unless {\it
all three of the following are satisfied}: a) The black hole is
maximally spinning, b) the black ring is maximally spinning ({\it
i.e.} it has zero entropy) and c) the charge of the black hole is
exactly the size needed for the black hole to only just capture the
black ring ($R \to 0$ as $\alpha
\to 0$). Hence, in the mergers with $\Delta S = 0$, the black ring must
itself have zero entropy, and it has to settle  on the equator of 
a maximally rotating black hole.

We therefore see that, at least for this sub-class of black holes and
black rings, we not only have chronology protection, but that the
merger process is, with one exception, thermodynamically irreversible.
The only reversible merger matches with physical intuition: One must
start with a black hole and a black ring both of zero entropy, and the
black ring must merge by just grazing the black hole equator.
While we have not analyzed the problem for the fully general three-charge
objects, we expect chronology protection to work in exactly the same way. 

Conversely, one can also investigate the condition for thermodynamically reversible 
mergers of fully-general black rings and black holes. In particular, suppose that
one starts with a zero-entropy black ring and a maximally-rotating black hole, 
both with arbitrary charges,  and one further imposes the physically sensible 
condition that the ring meets the black hole at the equator.  
For generic black holes and black rings we find that such a merger 
leads to a BMPV black hole of non-zero entropy. However, we also
find that there are thermodynamically reversible  mergers 
when the black ring and the black hole charges satisfy:
\eqn\grazing{\Nb_I  ~=~ {\Pb \over n^I}    \qquad {\rm and} \qquad
N^{BH}_I~=~   {P^{BH} \over n^I} \,, }
for two integers, $\Pb$ and $P^{BH}$.  Also note that this means 
\eqn\parallelcharges{N_I ~\equiv~ \Nb_I + \coeff{1}{2}\, C_{IJK} \,n^J\,n^K
~=~ {(\Pb +  n^1\,n^2 \,n^3) \over n^I}  \,,}
Thus the electric charges of black ring {\it and} its charges dissolved
in fluxes (${1 \over 2}\, C_{IJK} \,n^J\,n^K$) must both be aligned exactly
parallel to the electric charges of the black hole.

Our result suggests that the microstates corresponding to the maximally spinning 
BMPV black hole must belong to a special ensemble and that one cannot simply throw in any 
black ring microstate without rearranging the internal state of the BMPV black hole
and thereby generating entropy.  That is, to add a black ring microstate reversibly one has to
precisely prepare this microstate so as to match the ensemble to which it is being added.
We are presently investigating what this tells us about the ensemble of
microstates that make up the BMPV black hole.

\newsec{Conclusions and future directions}

We have constructed smooth, five-dimensional supergravity solutions
that describe a black ring and a black hole at an arbitrary distance
from the center of the black ring. We have found that as one moves the
black ring towards the black hole the angular momentum in the plane of
the ring remains constant, while the angular momentum in the plane
perpendicular to the ring increases.

We have also analyzed the merger of a black ring with a black
hole, and have shown that even if one starts with a solution where
the two angular momenta are different, to bring the black ring
into the black hole one has to change one of the angular momenta,
such that after the merger the two angular momenta are equal.

In \MarolfCX\ it was argued that by throwing a supertube into an
extremal (BPS) black hole one can create a non-extremal (non-BPS) 
black hole that is unstable. 
Our solutions show that {\it extremal} mergers (and presumably mergers that 
are very close to being extremal) can never 
produce an unstable object. Since supergravity solutions fully take 
into account back-reaction and angular momenta coming from fluxes, it 
would be very interesting to see whether one can use our solutions 
to learn more about the non-extremal mergers proposed in \MarolfCX.  
It would  also be worthwhile to try to extend our work to 
the non-extremal solutions, and thus obtain a complete description of non-extremal mergers.
A non-extremal extension of our solutions, even if only perturbative, will most certainly
illuminate the physics of the merger process and allow one to understand how 
the resulting unstable black hole evolves.

It is also possible to construct solutions where the black hole is
away from the center of the ring, but still in the plane of the
ring. Such a solution would be very interesting, because the
rotational invariance along the ring horizon would be broken (in
contrast to the solutions that we have constructed here). Hence, the 
near-ring solution will
probably look like the black rings with variable charge density
constructed in \BenaTD. If,  as claimed in \HorowitzJE, such solutions are
not smooth, we would have a process by which a smooth solution
would become singular as one moves on a moduli space - this would
be certainly very unexpected and interesting, and is currently under investigation.

Our configurations can be easily dualized to a frame in which the asymptotic 
charges are those of the D1-D5-P system. One can then take a near-horizon 
limit, and obtain an asymptotically $AdS_3 \times S^3 \times T^4$ solution, 
that contains a black hole and a black ring, and that is dual to 
an ensemble of boundary microstates. It would be very interesting to find 
what this ensemble is, and to give a microscopic description of the entropy 
of the black-ring black-hole configuration, similar to the microscopic 
description of the black rings in \BenaTK.
A way to attack this problem would be to compare the entropy of the 
ring-hole system to the entropy of a black ring with the same angular 
momenta and charges. If, for some values of the charges, the  single-ring 
entropy is smaller than that of a ring-hole system, then we would have a 
very interesting phase diagram, that will most definitely improve our 
microscopic understanding of these objects. Another route of attack would 
be to use the fact that one can get different angular momenta using the 
same ring and the same black hole placed at different distances. If there 
exists any link between the ring dipole charges and the length of string 
bits in the boundary dual \refs{\BenaTK,\ElvangDS,\AldayXJ\LuninIZ{--}\LuninJY}, then 
the fact then one obtains different angular momenta from the same objects
might help clarify their microscopic descriptions.

Another interesting aspect of the mergers we  have analyzed is the fact that a 
merger of a maximally spinning BMPV and a certain class of zero-entropy black rings 
appears to be thermodynamically {\it reversible}, unlike the merger of say 
two BMPV black holes. This indicates that if one adds a certain zero-entropy black ring 
microstate to  a  microstate correspond to
the maximally-spinning BMPV black hole, the result is another maximally spinning BMPV 
microstate.   Combining this with the knowledge of zero-entropy black ring 
microstates \refs{\BenaVA,\BerglundVB} might allow us to find the microstates 
corresponding to the maximally-spinning BMPV black hole, which would 
significantly improve our understanding of black holes in string theory.

\bigskip
\leftline{\bf Acknowledgments}

We would like to thank Don Marolf for helpful conversations and
 comments.  The work of NW and CWW is supported in part by the DOE
 grant DE-FG03-84ER-40168.  The work of IB is supported in part by the
 NSF grant PHY-0503584.

\appendix{A}{Gauge transformations of the angular momentum vector}

In our solution \kxres, the function $k_x$ diverges like ${1\over
\sqrt{1-x^2}}$ near the points $x=1$ and $x=-1$. Remembering that near
the horizon $x = - \cos \theta$, this implies that the rotation vector
$k_{\theta}$ is constant near $\theta=0$ and $\theta = \pi$. To put
the horizon metric in canonical form one can make a coordinate
transformation, which eliminates this constant.

If the function $k_x$ goes like ${-{a_1\over \sqrt{1-x}}}$ at the south
pole, and ${a_2\over \sqrt{1+x}}$ at the north pole, then the gauge
transformation
\eqn\gaugetr{ \vec k \rightarrow \vec k - \vec \nabla 
\left({(a_1(1+x) + a_2 (1-x)) \sqrt{1-x^2} \over \sqrt 2 }\right)}
eliminates the divergent parts of $k_{x}$ at the poles. 

By expanding $k_x$ we find that the function that enters the gauge
transformation is

\eqn\gfun{\eqalign{ G  ~\equiv~ & {D (1+x) \sqrt{1-x^2} \sin \phi \over  
2 R^2 \sqrt{1+\alpha^2}} 
\bigg[ {\log \left(\alpha ^2+1\right)\over 2 \alpha }-{\alpha  \over 2} 
\log \left({(y-1) \left(\alpha ^2+1\right) \over y \alpha ^2-\alpha^2+y+1}
\right) \cr
& \qquad \qquad  \qquad  \qquad  \qquad  \qquad 
-{2\alpha  \over y \alpha^2-\alpha^2+y+1} \bigg]  \cr 
&+ {D (1-x) \sqrt{1-x^2} \sin \phi \over  
4 \alpha R^2 \sqrt{1+\alpha^2}} 
\left[ \log \left(\alpha ^2+1\right)-
\log \left({(y-1) \left(\alpha ^2+1\right) \over y
   \alpha ^2+\alpha ^2+y-1}\right)\right]
}}

and the new components of the angular momentum vector will be
\eqn\newk{\eqalign{k_x' &= k_x - \partial_x G \,, \qquad
k_y'  =  - \partial_y G \cr
k_\psi' &= k_\psi  \,, \qquad
k_\phi'  = k_\phi - \partial_\phi G~.  
 }}

\vfill\eject
\listrefs
\vfill\eject

\end